\documentclass[]{spie}  %>>> use for US letter paper
\usepackage{amsmath,amsfonts,amssymb}
\usepackage{graphicx}
\usepackage[colorlinks=true, allcolors=blue]{hyperref}

\title{Performance of attention-based deep-learning models in Shack-Hartmann wavefront sensor measurements}

\author[a, b]{Matias Marambio-Jimenez}
\author[a, b]{Yosuke Minowa}
\author[b]{Yoshito Ono}
\author[a, b]{Hajime Ogane}
\author[a, b]{Shin Oya}
\author[a, b]{Yusei Koyama}
\author[c]{Masayuki Akiyama}
\affil[a]{Astronomical Science Program, Graduate University for Advanced Studies, SOKENDAI, 2-21-1 Osawa, Mitaka, Tokyo 181-8588, Japan}
\affil[b]{National Astronomical Observatory of Japan, 2-21-1 Osawa, Mitaka, Tokyo 181-8588, Japan}
\affil[c]{Astronomical Institute, Tohoku University, 6-3 Aramaki, Aoba, Sendai 980-8578, Japan}

\authorinfo{Further author information: (Send correspondence to M.M.J)\\M.M.J.: E-mail: matias.marambio@grad.nao.ac.jp}

\begin{document} 
\maketitle

\begin{abstract}

A systematic performance evaluation of attention-based deep-learning models for wavefront reconstruction in Shack-Hartmann wavefront sensor (SHWFS) systems is presented. While deep learning has been extensively applied to pyramid and focal-plane wavefront sensing, its application to standard SHWFS configurations remains relatively underexplored. A simulation dataset consisting of 3.5 million open-loop uncorrelated turbulent phase and image pairs modeled after the Subaru Telescope pupil was generated. The dataset spans four pupil-sampling cases (8$\times$8, 16$\times$16, 20$\times$20, and 32$\times$32 subapertures) under realistic conditions. A custom Vision Transformer (ViT) which incorporates features such as rotary positional embeddings, register tokens, and a cross-attention decoder, is benchmarked against a fully convolutional U-Net and a classical Maximum A Posteriori (MAP) linear reconstructor. The results show that the ViT architecture yields the lowest overall residual phase variance across all tested subaperture configurations, demonstrating a robust capability to model global spatial correlations and spot-to-spot cross-coupling across the pupil. While the U-Net performs competitively by capturing localized features, it exhibits $\sim1.5$ times more residual variance per mode than the ViT, which is relatively significant in high-order modes. This study represents a key step toward implementing attention-based models for high-order wavefront sensing and telemetry-assisted data reduction in next-generation instrument suites like ULTIMATE-Subaru.

\end{abstract}

% Include a list of up to six keywords after the abstract
\keywords{Deep learning for Shack-Hartmann wavefront sensors, Vision Transformers, Subaru telescope dataset for machine learning in adaptive optics, Deep learning applied to adaptive optics, U-Net}

\section{Introduction}
\label{sec:intro}  % \label{} allows reference to this section
Recent work in adaptive optics (AO) has explored deep learning using reinforcement learning\cite{2022OExpr..30.2991P, rl1,PO4AO1, PO4AO2, PO4AO3} and transformer-based models\cite{ViTinPyr, ViTinFP}, but almost all have been focused on pyramid wavefront sensors\cite{2021JATIS...7a9001W, 2023PASP..135k4501W, landman1, landman2} or focal plane wavefront sensing \cite{jalofp, fp2}, leaving conventional Shack-Hartmann wavefront sensor (SHWFS) based systems underexplored in this context. Nevertheless, it has been shown that deep-learning models are capable of interpreting full-frame image information from SHWFS frames to reconstruct high-frequency components of the turbulent phase\cite{2023JATIS...9a9001S}.

\subsection{Self-attention}
Traditional wavefront reconstruction methods for SHWFS rely on localized, independent centroid measurements, which inherently ignore long-range spatial correlations and spot-to-spot cross-coupling across the pupil. To overcome these kind of limitations in different data modalities, recent advancements in deep learning have introduced the now widely used attention mechanism, originally proposed by Vaswani et al.~\cite{vaswani}. Rather than processing localized subaperture images through translation-invariant convolutional filters, like in convolutional neural networks (CNNs), attention-based models, like vision transformers (ViTs)\cite{ViT} partition the SHWFS frame into a sequence of flattened, tokenized patches that naturally align with the grid of subapertures. This allows the network to dynamically compute pairwise affinities between all subapertures across the entire pupil simultaneously. The core of this mechanism is the scaled dot-product attention, mathematically formulated as:

\begin{equation}\label{eq:attention}
    \text{Attention}(\mathbf{Q}, \mathbf{K}, \mathbf{V}) = \text{softmax}\left(\frac{\mathbf{Q}\mathbf{K}^T}{\sqrt{d_k}}\right)\mathbf{V},
\end{equation}

where $\mathbf{Q}$, $\mathbf{K}$, and $\mathbf{V}$ are the query, key, and value projection matrices of the patch embeddings (subaperture embeddings), respectively, and $d_k$ is the dimension of the keys. 

In the physical context of a SHWFS frame, this formulation provides a powerful framework for multi-scale wavefront sensing. In self-attention, the queries $\mathbf{Q}$ and keys $\mathbf{K}$ project the localized spatial features of each subaperture, such as spot shape, intensity and local displacement, to compute a global affinity matrix. The dot product $\mathbf{Q}\mathbf{K}^T$ represents the statistical and spatial correlation between different regions of the pupil. 
%When a spot experiences non-linear distortion, such as truncation near the primary mirror boundaries or the asymmetric spider supports, the attention mechanism dynamically allocates higher weights to surrounding, well-behaved subapertures. 
This enables the model to leverage global context to resolve local ambiguities. Finally, the value matrix $\mathbf{V}$ encodes the physical phase information within each subaperture, which is weighted and aggregated by the softmax attention map. By doing so, the ViT naturally incorporates spatial and boundary priors directly from the raw intensity distribution, leading to a highly robust reconstruction of both low-order aberrations and high-frequency phase details.

\subsection{Cross-attention}
While self-attention maps the internal relationships within the raw SHWFS intensity frame, the concept can be generalized to two different domains through the implementation of \textit{cross-attention} (or encoder-decoder attention)~\cite{vaswani}. In an encoder-decoder architecture , cross-attention acts as the mathematical bridge between the source wavefront sensor measurements and the target phase reconstruction space. Here, the encoder processes the raw 2D SHWFS frame to generate a set of key ($\mathbf{K}_{\text{enc}}$) and value ($\mathbf{V}_{\text{enc}}$) representations that capture the localized spot geometries, intensity distributions, and noise characteristics. Meanwhile, the decoder, which is responsible for synthesizing the target 2D phase map, generates queries ($\mathbf{Q}_{\text{dec}}$) representing the target phase grid pupil information, like coordinates and throughput. The cross-attention operation then evaluates:

\begin{equation}\label{eq:cross_attention}
    \text{Cross-Attention}(\mathbf{Q}_{\text{dec}}, \mathbf{K}_{\text{enc}}, \mathbf{V}_{\text{enc}}) = \text{softmax}\left(\frac{\mathbf{Q}_{\text{dec}}\mathbf{K}_{\text{enc}}^T}{\sqrt{d_k}}\right)\mathbf{V}_{\text{enc}}.
\end{equation}

Physically, this allows the target reconstruction space to dynamically ``query'' the raw WFS image features at the exact spatial locations and subapertures containing the relevant gradient information. This explicit cross-domain mapping provides a highly flexible and physically interpretable mechanism for translating raw intensity distributions into direct phase maps, bypassing the rigid linear assumptions of classical geometric reconstructors.

In this paper, an AO post-processing system that applies attention-based ViTs directly to open-loop SHWFS images was developed and benchmarked against CNNs in the form of U-Net\cite{U-Net} models and Maximum A Posteriori (MAP) reconstructors.
CNNs and ViTs differ in their architectural inductive biases: CNNs are hierarchical and convolutional, favoring local, translation-invariant features, whereas ViTs use attention layers that model long-range dependencies with minimal built-in assumptions, guided only by positional encodings. This work systematically compares these paradigms for SHWFS data, testing whether attention mechanisms better capture complex, non-local aberration patterns relevant to high-order correction.

The Subaru telescope provides a unique environment to build standardized, publicly usable datasets from the AO3k\cite{AO3k} and ULTIMATE-START\cite{start1, start2} Laser Tomography AO (LTAO) system, which enables AO in visible wavelength for 8m-class telescopes. This study is a first-step on building an attention-based deep-learning model for AO telemetry assisted data reduction in AO systems in the Subaru telescope. The developed reduction dataflow can be utilized in the next-generation AO system including ULTIMATE-Subaru\cite{ultimate}, a wide-field near-infrared instrument with Ground Layer AO (GLAO).

This paper is organized as follows. First, dataset generation is explained in detail in section \ref{sec:dataset}. Next, the model training method and the architecture of each model is explained in section \ref{sec:training}. The results of the model comparison are shown in section \ref{sec:comparison}. Finally, the summary and future work are presented in section \ref{sec:summary} and acknowledgments in section \ref{sec:acknowledgements}.

\section{Dataset Generation}\label{sec:dataset}

A dataset that represents the real case-scenario of the ULTIMATE-START was chosen to evaluate different wavefront reconstruction models. As a first proof of concept, four different simulated open-loop natural guide star datasets were generated with a modified version of \texttt{OOPAO}\cite{oopao} that can generate dataset samples in parallel batches using \texttt{pytorch}\cite{pytorch}. Since the objective is to characterize the error budget of the different wavefront reconstructors, it was decided to base the dataset on different sampling cases of the telescope pupil. The lowest reconstruction variance case (highest measurement variance) is a 32$\times$32 Shack-Hartmann wavefront sensor based on the truth-WFS from the ULTIMATE-START LTAO system, while the highest reconstruction variance case (lowest measurement variance) is a SHWFS of the same characteristics but with only 8$\times$8 subapertures over the telescope pupil. The two remaining datasets correspond to the intermediate cases of 16$\times$16 and 20$\times$20 subapertures over the telescope pupil. To simplify the error budget analysis, temporal effects on the power spectral density of the phase due to integration time of the SHWFS were not included in the simulation. 

The parameters of the simulation are presented in table \ref{tab:simparams}. The main experiment at this point of the research was to evaluate the models performance on constant turbulence characteristics but with varying noise, to validate that a ViT can be trained on a simpler case. The median turbulence characteristics of the latest site testing were used for this simulation\cite{atmosphericparameters}. The parameters that were varied randomly in the simulation (i.e. signal photon count, background surface brightness, and RON) were sampled from a uniform distribution. This makes the dataset represent extreme cases as strongly as nominal cases, which can help a deep-learning model generalize to these situations.

\begin{table}[b]
\caption{Simulation parameters used for the Shack-Hartmann wavefront sensor frame and atmospheric phase simulation. Parameters in between brackets are sampled from a uniform distribution. Atmospheric characteristics are from the TMT site located 90 m below the Subaru site\cite{atmosphericparameters}. } 
\label{tab:simparams}
\begin{center}       
\begin{tabular}{|l|l|} %% this creates two columns
%% |l|l| to left justify each column entry
%% |c|c| to center each column entry
%% use of \rule[]{}{} below opens up each row
\hline
\rule[-1ex]{0pt}{3.5ex}  Parameter & Value  \\
\hline\hline
\rule[-1ex]{0pt}{3.5ex}  Pupil &\, Subaru Telescope (built with \texttt{poppy}\cite{poppy})    \\
\hline
\rule[-1ex]{0pt}{3.5ex}  Subaperture grid size &\, 32$\times$32, 20$\times$20, 16$\times$16, 8$\times$8  \\ 
\hline
\rule[-1ex]{0pt}{3.5ex}  Pixels per subaperture &\, 8$\times$8 pix\\
\hline
\rule[-1ex]{0pt}{3.5ex}  Phase grid size &\, 320x320   \\
\hline
\rule[-1ex]{0pt}{3.5ex}  Fried parameter (500 nm) &\, 0.118 m   \\
\hline
\rule[-1ex]{0pt}{3.5ex}  Outer scale &\, 20 m   \\
\hline
\rule[-1ex]{0pt}{3.5ex}  Effective wavelength &\, 658 nm   \\
\hline
\rule[-1ex]{0pt}{3.5ex}  Pixel scale &\, $0.548"/\mathrm{pix}$  \\
\hline
\rule[-1ex]{0pt}{3.5ex}  Signal photon count &\, $[\sim100,\,\sim2000]$ ph \\
\hline
\rule[-1ex]{0pt}{3.5ex} Background surface brightness &\, $[5\times10^3,\,5\times10^4]$ ph/s/arcsec$^2$ \\
\hline
\rule[-1ex]{0pt}{3.5ex} RON &\, $[0.7, 2.0]$ e-/pix  \\
\hline
\rule[-1ex]{0pt}{3.5ex}  Sensor bit depth &\, 16 bits  \\
\hline
\rule[-1ex]{0pt}{3.5ex}  Sensor quantum efficiency & \, 0.8 \\
\hline
\rule[-1ex]{0pt}{3.5ex}  Sensor full well capacity & \, 30000 e- \\
\hline 
\rule[-1ex]{0pt}{3.5ex}  Exposure time &\, 2 ms   \\
\hline
\end{tabular}
\end{center}
\end{table} 

The simulation includes two versions of the SHWFS frame, a star-signal only version and a full version including background signal, photon noise, and detector noise. The dataset triplet (phase, signal-only frame, full frame) data are saved as a shards using \texttt{WebDataset}\cite{webdataset}, with the phase saved as OPD from flat wavefront in microns at the SHWFS effective wavelength. Each dataset is split into a training, validation and test datasets, containing 80\%, 10\%, 10\% of the data respectively. The mean and standard deviation of the full frames of the training splits of the datasets are saved for standardization during data loading into the deep-learning models.

\section{Model training}\label{sec:training}

Two different 3-million parameter deep-learning models were developed for benchmarking their performance: a U-Net and a ViT. The ViT models' hidden layers were trained with the Muon optimizer\cite{muon1} using the matching root-mean squared (RMS) AdamW approach\cite{muon2}, while the remaining layers and the U-Net models were trained using the AdamW\cite{adamw} optimizer. The models were trained using the mean squared error (MSE) of the residuals over the valid pupil phase pixels.

To train these two models in a fair manner, an optimizer hyperparameter search was carried out over each dataset using low-discrepancy quasi-random random sequences, specifically a scrambled Sobol sequence\cite{sobol1, sobol2, sobol3, sobol4}. The hyperparameters tuned are learning rate, momentum ($\beta_1$), and weight decay using 30 draws. To establish the learning rate range for the draws, a training run with MSE loss until divergence was ran with increasing learning rate per step for each dataset and model. The range was chosen by visually studying the validation loss curve. For each draw, a small scale 10,000 steps training run was carried out, choosing the best performing combination for further training. All of the training was done in a single A100 80GB GPU.

The chosen models for each dataset were trained up to 60,000 steps, using a learning rate schedule of 2,500 linear warm-up steps, cosine annealing down to a minimum learning rate of $10^{-6}$, and 5,000 linear cool-down steps to a learning rate of $10^{-12}$ with gradient clipping to stabilize training.  There were some cases for the ViT were the best performing model in the small-scale hyperparameter search diverged when trained further. In those cases, the next best performing model was trained, resolving the issue. Training of the ViTs using only the AdamW optimizer was also investigated; but the models diverged, suggesting a very complicated loss landscape.

\begin{figure}[t]
   \begin{center}
   \begin{tabular}{c} %% tabular useful for creating an array of images 
   \includegraphics[width=\linewidth]{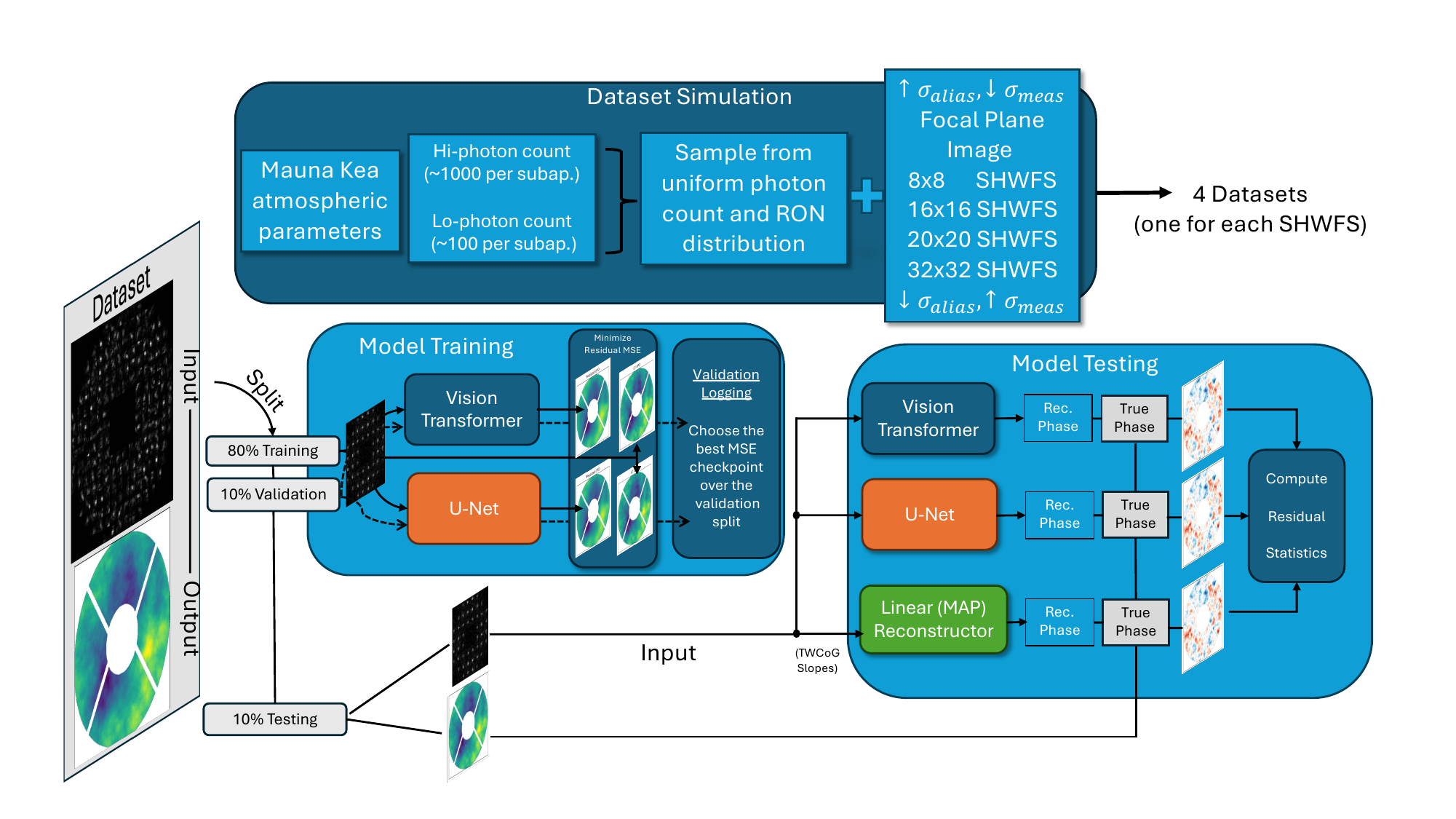}
   \end{tabular}
   \end{center}
   \caption[methods] 
%>>>> use \label inside caption to get Fig. number with \ref{}
   { \label{fig:researchmethods} 
Diagram of the research methods. The top part shows the dataset generation characteristics. The bottom part shows how the models are trained and evaluated.}
   \end{figure} 

\subsection{U-Net}\label{sec:unet}

\begin{figure}[t]
   \begin{center}
   \begin{tabular}{c} %% tabular useful for creating an array of images 
   \includegraphics[width=\linewidth]{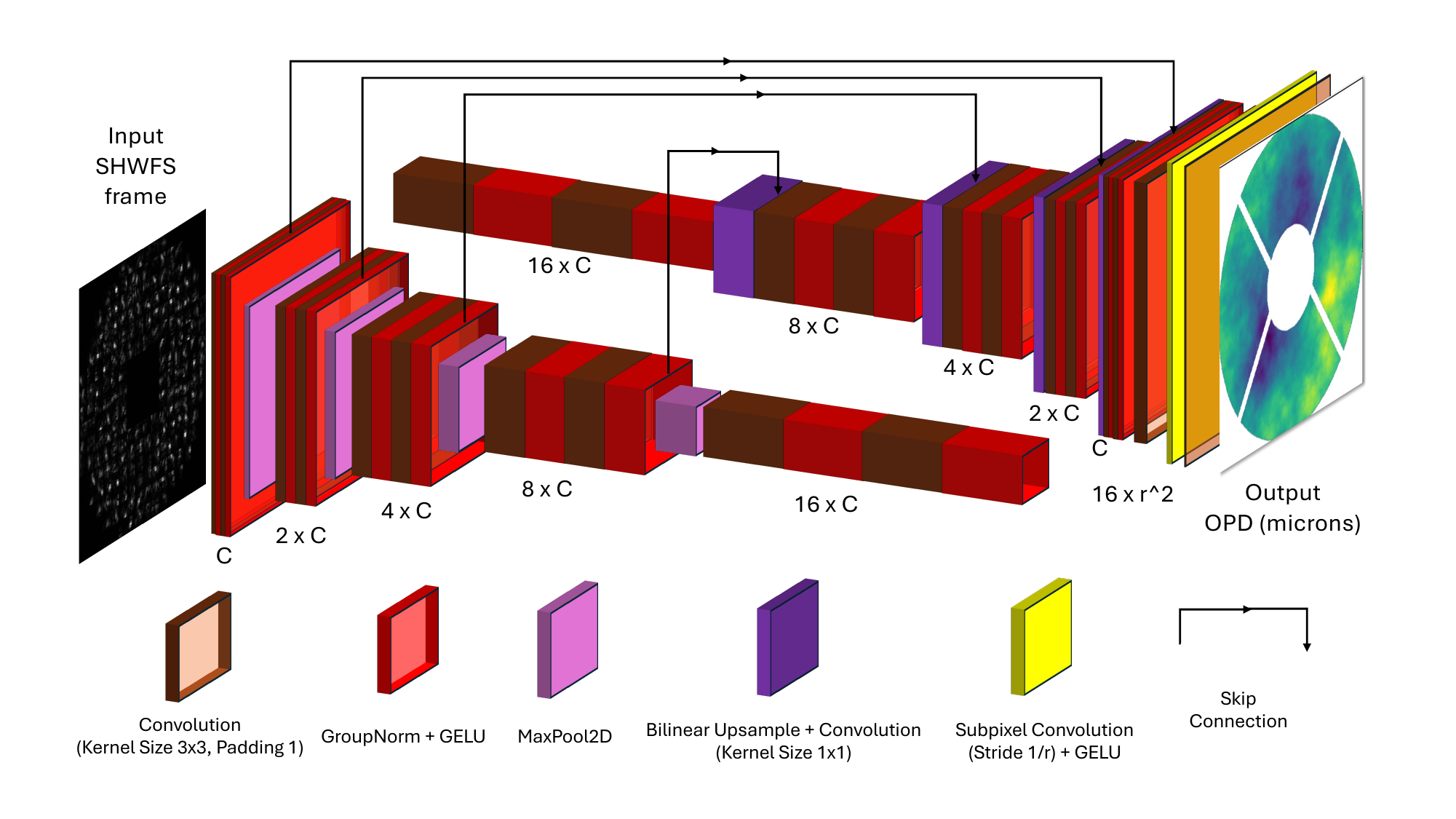}
   \end{tabular}
   \end{center}
   \caption[unet] 
%>>>> use \label inside caption to get Fig. number with \ref{}
   { \label{fig:unet} 
        Diagram of the U-Net architecture. Each MaxPool2D\cite{maxpool2d} down-sample layer output is added with a convolutional layer to achieve linear residual connections. Furthermore, each encoder block output which has the same number of channels as its decoder counterpart gets added to the decoder block input as a skip connection. $C=18$ corresponds to the length of the base channel dimension in the U-Net. $r$ is the upscaling factor for the subpixel convolution. For each dataset $r=5$, except for the $20\times20$ SHWFS dataset, where $r=2$. The model uses GroupNorm\cite{groupnorm} and Gaussian Error Linear Unit (GELU) activation functions \cite{gelu}.}
\end{figure}

The U-Net built in this study is based on the Deep Residual U-Net\cite{resunet}. This model is a fully convolutional encoder-decoder network characterized by a symmetric contracting path (encoder) and an expanding path (decoder) joined by skip connections. The encoder extracts multi-scale spatial features through successive convolutional layers and down-sampling, while the decoder up-samples the resulting feature maps from the encoder to project the localized spatial representation to phase space. The skip connections preserve and concatenate high-resolution details directly from the contracting path to the expanding path, as shown in figure \ref{fig:unet},
preventing the loss of high-frequency information. However, the U-Net decoder output size is limited to the input size, so up-sampling is necessary for the output to be of size 320x320. Each consecutive U-Net block is further connected by a linear convolutional layer, realizing a residual connection between each down-scaling and up-sampling layer. The final layers (phase head) of the decoder consist of an up-sampling layer which restore the spatial dimensions of the compressed feature maps using a learnable sub-pixel convolution\cite{subpixelconv} with a GELU activation function, instead of fixed interpolation layers. It is followed by a 2D convolution which maps the 16 output channels of the subpixel convolution layer to phase space in OPD (microns).

\subsection{Vision Transformer}\label{sec:vit}

The ViT implemented in this study is based on a collection of recent advances on ViT design, including LayerScale\cite{layerscale}, RMSNorm\cite{rmsnorm}, Rotary Positional Embedding (RoPE)\cite{rope}, register tokens\cite{registers} and QK-norm\cite{qknorm}. It has been shown that the combination of these modifications result in better performance in both classification and regression tasks\cite{vit-5}. 

\begin{figure}[t]
   \begin{center}
   \begin{tabular}{c} %% tabular useful for creating an array of images 
   \includegraphics[width=\linewidth]{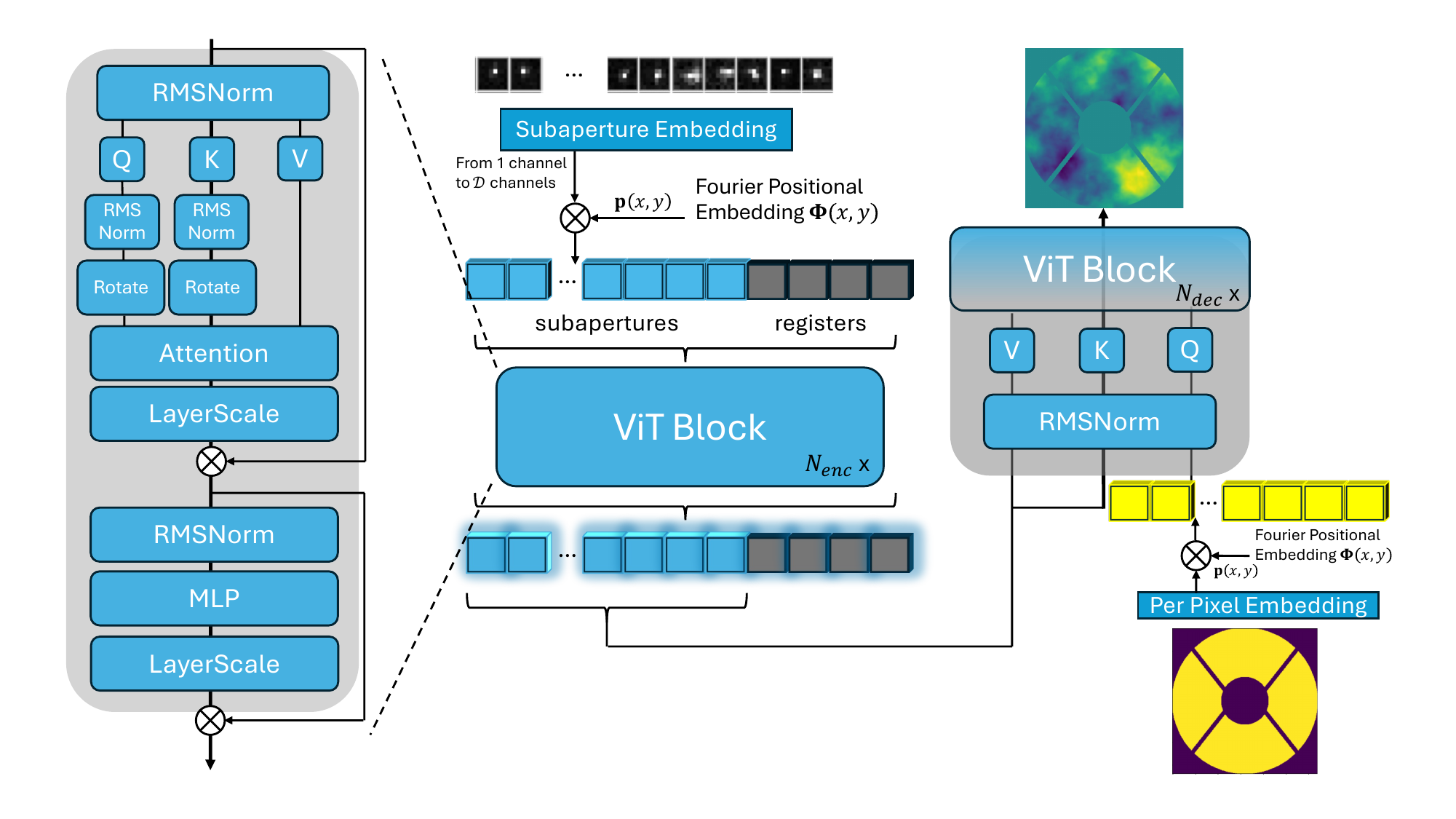}
   \end{tabular}
   \end{center}
   \caption[vit] 
%>>>> use \label inside caption to get Fig. number with \ref{}
   { \label{fig:vit} 
        Diagram of the ViT architecture. To the left the internals of a ViT block are shown, with data flowing top-down. In the middle, an overall view of the ViT encoder is shown, with subaperture tokens shown in blue, and register tokens in gray. To the right, an overall view of the ViT decoder is shown, showing the query tokens in yellow.}
\end{figure} 

The ViT model, shown in the diagram of figure \ref{fig:vit}, processes the raw SHWFS frame by dividing the input image into a grid of non-overlapping patches corresponding to the subapertures of the SHWFS. These patches are linearly projected into subaperture-embeddings, and summed element-wise with absolute 2D Fourier positional embeddings in pupil space. To construct these positional embeddings, the physical subaperture grid centers are mapped onto a normalized spatial interval $[-1, 1]\times[-1,1]$, where the boundaries align with the pupil edges in phase space. To try to avoid the introduction of aliasing, a series of logarithmically spaced (dyadic) frequency bands are built $f_k=2^k,$ for $k=0, 1, 2,...$, where the maximum frequency is bounded by the grid's Nyquist limit.  For each subaperture center, the Fourier positional embedding, $\mathbf{\Phi}(x,y)$ on the normalized domain $\{(x,y)\in\mathbb{R}^2|-1\leq x\leq1,\,-1\leq y\leq1\}$ is generated by projecting the coordinates through sine and cosine functions at each frequency band $f_k$, generating a $4k$-dimensional vector (consisting of $2k$ dimensions for the horizontal coordinate and $2k$ dimensions for the vertical coordinate) . This embedding is then projected to the transformer hidden dimension $\mathcal{D}$ using a learnable linear layer:
\begin{equation}\label{eq:linearproj}
    \mathbf{p}(x,y)=g\cdot(\mathbf{W_{proj}\Phi}(x,y)+\mathbf{b}),
\end{equation}
where $g$ is a learnable gain parameter, $\mathbf{W_{proj}}$ is the learnable weight tensor, $\mathbf{b}$ is a learnable bias, and $\mathbf{p}(x,y)$ is the projection of the position embedding onto the $\mathcal{D}$ dimensions.  

Summarizing, to produce the resulting tokens, a learnable patch projection layer maps each subaperture patch from an $8\times8$ single-channel image to a $\mathcal{D}$-dimensional vector. The $4k$-dimensional positional embedding vector is projected to $\mathcal{D}$-dimensions using equation \eqref{eq:linearproj} and added directly to the subaperture embedding to produce the subaperture tokens.  Before routing the resulting token sequence into the transformer encoder blocks, 4 learnable register tokens are concatenated to the sequence to help absorb high-frequency attention outliers (attention sinks)\cite{attentionsink}.

The combined sequence of subaperture tokens and register tokens pass through 12 consecutive ViT blocks. In each block, the tokens are first normalized using RMSNorm and then individually projected into query key and value with three different learnable linear layers. Before entering the multi-head self attention layer, composed of 6 parallel attention heads, the query and key vectors are normalized one more time (QK-norm) using RMSNorm. Additionally, relative-position-dependent feature matching is enforced using 2D RoPE. The attention head dimension is bisected into horizontal and vertical components, and independent 1D RoPE rotations are applied. This separation dedicates the first half of the attention head dimensions entirely to encoding horizontal spatial relationships and the second half to vertical spatial relationships.

The rotation angles are computed by scaling the normalized coordinate positions with a set of inverse frequencies
\begin{equation}
    \theta_{n,d}=p_n\cdot w_d,
\end{equation}
where $p_n$ is the coordinate ($x$ or $y$) of token $n$, $w_d=\mathrm{base}^{-2d/\mathrm{half}}$ is the inverse frequency of the $d$-th pair of dimensions, where $\mathrm{base}=10^{4}$ for subaperture tokens and $\mathrm{base}=10^{3}$ for register tokens. Since register tokens do not correspond to any physical coordinate on the pupil,  they are assigned to a virtual grid in a $2\times2$ coordinate system for the 4 registers. 

Following the self-attention layer, a residual connection modified by a learnable diagonal scaling vector (LayerScale) is applied to generate intermediate features. These intermediate features are then normalized using RMSNorm and passed to a multi-layer perceptron (MLP)\cite{welch2025welch} with a hidden dimension expansion ratio of 4. A second LayerScale scaling and residual connection produce the final output of the block. The output of these 12 ViT blocks is a rich representation of the SHWFS frame that gets fed into the ViT decoder.

The decoder constructs a dense set of query representations designed to map directly onto the wavefront phase. The model takes a 2D  pupil mask, flattens it to a 1D sequence of pupil throughput values, projects them into the ViT hidden dimension $\mathcal{D}$, and adds the corresponding 2D pupil-space Fourier positional embeddings projected via equation \eqref{eq:linearproj}. This sequence of target pixel query tokens is input into 3 consecutive cross-attention ViT blocks also with 6 attention heads each. 

These decoder blocks share the same structure as the encoder blocks. However, the attention steps compute cross-attention instead of self-attention. Specifically, the queries $\mathbf{Q}_{\text{dec}}$ are projected from the pupil-pixel throughput tokens, while the keys $\mathbf{K}_{\text{enc}}$ and values $\mathbf{V}_{\text{enc}}$ are projected from the subaperture tokens output by the encoder (discarding the register tokens). This formulation enables each phase pixel to selectively attend to relevant spatial regions of the encoded SHWFS frame, effectively executing a dynamic, spatially aware wavefront reconstruction. No register tokens are added to the query tokens sequence. 

At this point the reconstructed phase is of shape $[\mathcal{B}, 102400, \mathcal{D}]$, with $\mathcal{B}$ corresponding to the batch size. This output from the decoder ViT blocks is projected through a final linear layer to reduce the number of channels from $\mathcal{D}$ to 1, corresponding to the phase OPD value in microns.  Finally, this sequence is transposed and reshaped back into a 2D spatial grid, yielding the reconstructed phase map of shape $[\mathcal{B}, 1, 320, 320]$. For this experiment, $\mathcal{D}=144$, and all the activation layers of the ViT are GELUs.

\subsection{Linear Reconstructor}\label{sec:linear}

To benchmark the performance of the deep-learning models, a Maximum A Posteriori (MAP) linear reconstructor was implemented described by
\begin{equation}
\left(\mathbf{G^T C_\eta^{-1}G+C_a^{-1}}\right)\mathbf{a}=\mathbf{G^T C_\eta^{-1}s},
\end{equation}
where $\mathbf{G}$ is the Southwell geometry matrix\cite{southwell}, $\mathbf{C_a}=\mathrm{diag}(\lambda_1,\dots,\lambda_M)$ is the diagonal covariance matrix prior of $M$ Karhunen-Loève (KL) variances, $\mathbf{C_\eta}$ is the dynamic noise covariance matrix from centroid estimation\cite{centroids}, $\mathbf{a}$ is the KL mode coefficient vector, and $\mathbf{s}$ is the measured slope vector before any regularization.

$\mathbf{G}$ is calibrated using $20,000$ SHWFS frame-phase pairs from the training split of the dataset using a weighted least squares formulation. For a slope measurement $i$ over $K$ calibration samples:
\begin{equation}
    \min_{\mathbf{G}_i}\sum_{k=1}^K \mathbf{W}_i(k)(\mathbf{s}_i(k)-\mathbf{G}_i\mathbf{a}(k))^2,
\end{equation}
where $\mathbf{W}_i(k)=1/\sigma_{ik}^2$ is the noise variance for the $i$-th slope in the $k$-th sample.

A covariance matrix for Thresholded-Weighted Center of Gravity (TWCoG) centroiding method was implemented for this linear reconstructor. This estimator defines the subaperture centroid as:
\begin{equation}
    \mathbf{c}_T=\dfrac{\sum_{k\in S_T}w_k(I_k-T_k)\mathbf{r}_k}{\sum_{k\in S_T}w_k(E[I_k]-T_k)},
\end{equation}
where $\mathbf{r}_k=[x_k, y_k]^T$ represents the pixel coordinate grid inside the subaperture window, $w_k=I_{\mathrm{ref},k}$ is the static spatial template weight of pixel $k$, $I_k$ is the measured pixel intensity, $T_k=3\sigma_{\mathrm{RON}}+0.05\max(I_k)$ is the intensity threshold where $\sigma_{\mathrm{RON}}=1.6$e-/pix, and $S_T=\{k\,|\,E[I_k]\geq T_k\}$ is the subset of pixels whose expected intensity ($E[I_k]=N_{\mathrm{ph}}I_{\mathrm{ref},k}$) exceeds the threshold.

Assuming independent pixel noise, a first-order multivariate Taylor expansion is used to calculate the covariance matrix of the centroid estimator:
\begin{equation}\label{eq:covthomas}
    \mathbf{C}_{\eta, \mathbf{c}_T}\approx\dfrac{\sum_{k\in S_T}w_k^2\sigma_k^2\mathbf{r}_k\mathbf{r}_k^T}{\left(\sum_{k\in S_T}w_k(E[I_k]-T_k)\right)^2},
\end{equation}
where $\sigma_k^2=E[I_k]+\sigma^2_{\mathrm{RON}}$.

It is common to include a slope response coefficient $\gamma_s$ to maintain a linear measurement scale. It can be defined as:
\begin{equation}
    \gamma_s=\dfrac{\sigma_{\mathrm{temp}}^2}{\sigma_{\mathrm{temp}}^2+\sigma_{\mathrm{spot}}^2},
\end{equation}
where $\sigma_{\mathrm{temp}}^2$ is the spatial variance (width) of the template spot, which is much broader than the short exposure instantaneous spot, and $\sigma_{\mathrm{spot}}^2$ is the spatial variance of the instantaneous spot. Since the reference template is constructed by taking the pixel-wise average frame over $20,000$ temporally uncorrelated dataset frames, it represents a long-exposure (infinite) template rather than a short-exposure template, indicating a $\gamma_s=1$.

During experiments, it was noted that since the simulated dataset is well-behaved, the spots in each subaperture do not transition into an extremely low signal-to-noise ratio (SNR) regime. Therefore, in addition to studying the MAP reconstructor with the analytical centroid measurement covariance defined in equation \eqref{eq:covthomas}, the Crámer-Rao Lower Bound (CRLB)\cite{covcrlb} case, described by equation \eqref{eq:covcrlb}, was also evaluated. The CRLB represents the theoretical limit of centroiding precision for an unbiased statistical estimator\cite{centroids}, and its covariance matrix is defined as:
\begin{equation}\label{eq:covcrlb}
    \mathbf{C}_{\eta, \text{CRLB}} = \frac{1}{N_{\text{ph}}^2} \left[ \sum_{k=1}^{N_{\text{pix}}} \frac{\mathbf{D}_k \mathbf{D}_k^T}{\sigma_k^2} \right]^{-1},
\end{equation}
where the vector $\mathbf{D}_k$ represents the spatial gradient of the reference template at pixel $k$, defined as:
\begin{equation}
    \mathbf{D}_k = \begin{bmatrix} -\frac{\partial I_{\text{ref}, k}}{\partial x} \\[6pt] -\frac{\partial I_{\text{ref}, k}}{\partial y} \end{bmatrix}.
\end{equation}
\begin{figure}[t]
   \begin{center}
   \begin{tabular}{c} %% tabular useful for creating an array of images 
   \includegraphics[width=\linewidth]{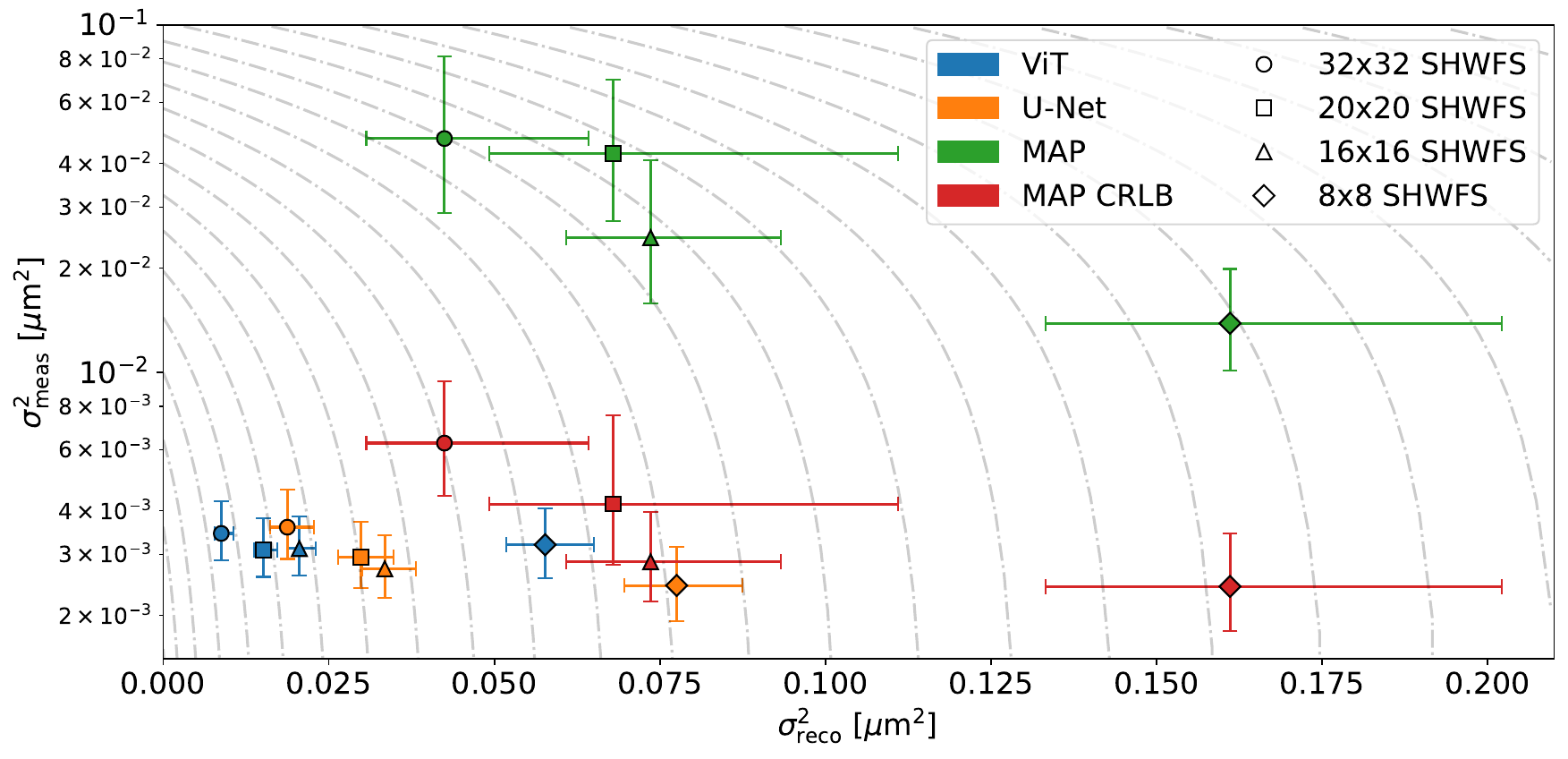}
   \end{tabular}
   \end{center}
   \caption[errorbudget] 
%>>>> use \label inside caption to get Fig. number with \ref{}
   { \label{fig:errorbudget} 
    Median of the variance components of each of the models evaluated for each testing dataset. Dotted lines correspond to constant total variance curves. Circles represent the 32$\times$32 SHWFS dataset, squares represent the 20$\times$20 SHWFS dataset, triangles represent the 16$\times$16 SHWFS dataset, and diamonds represent the 8$\times$8 SHWFS dataset. Error bars are calculated on 1 standard deviation over the entire testing dataset.}
\end{figure} 
For the KL mode prior, the first 1000 KL modes excluding piston were computed. Reconstruction was also carried out using these KL modes.

\section{Model comparison}\label{sec:comparison}

\begin{figure}[t]
   \begin{center}
   \begin{tabular}{c} %% tabular useful for creating an array of images 
   \includegraphics[width=\linewidth]{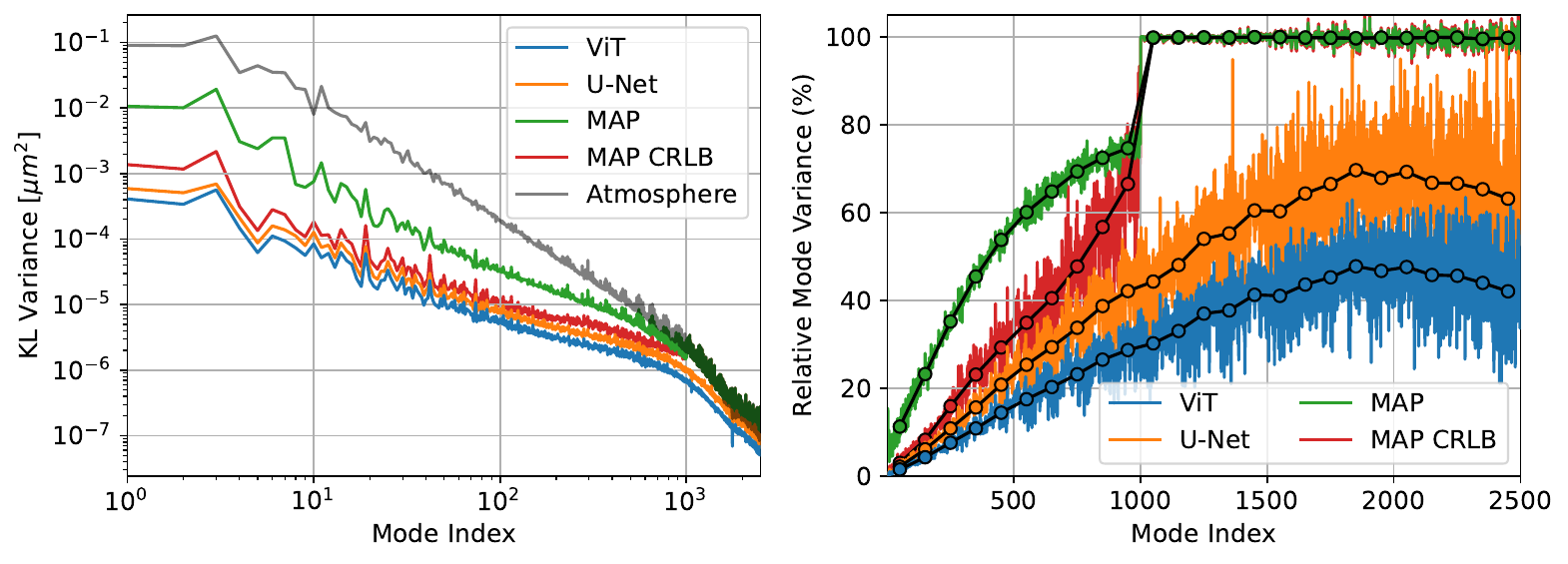}
   \end{tabular}
   \end{center}
   \caption[klvariance] 
%>>>> use \label inside caption to get Fig. number with \ref{}
   { \label{fig:klvariance} 
    \textbf{Left:} Median of the residual modal variance over the 32$\times$32 SHWFS testing dataset in addition to the ground-truth atmospheric phase. $2,500$ KL modes were used to fit the residual between the ground-truth phase and the model output. The blue line corresponds to the ViT reconstructor, the orange line to the U-Net reconstructor, the green line to the analytical covariance MAP reconstructor, the red line to the CRLB covariance MAP reconstructor, and the transparent black line to the ground-truth atmospheric phase. \textbf{Right:} Median residual variance relative to total ground-truth atmospheric variance for each KL mode. $2,500$ KL modes were used to fit the phase. Same color as the left plot for each reconstructor are used. The corresponding binned data-points every 100 modes are overlaid over each curve for clarity.}
\end{figure} 
To separate the variance contributions of the different error sources, it was assumed each error source is independent from the others:
\begin{equation}\label{eq:errorbudget}
\sigma^2=\sigma_{\mathrm{meas}}^2+\sigma_{\mathrm{reco}}^2,
\end{equation}
where $\sigma^2$ is the total variance, $\sigma_{\mathrm{meas}}^2$ is the variance attributed to all the noise present in the SHWFS frame, $\sigma_{\mathrm{reco}}^2$ is the reconstruction variance and it is defined differently for MAP reconstructors and neural network reconstructors. In the case of MAP reconstructors, $\sigma_{\mathrm{reco}}^2$ is the combination of errors attributed to aliasing and fitting error. However, since there is no formal definition of a fitting error or aliasing error in the case neural networks, $\sigma_{\mathrm{reco}}^2$ in this case is defined as the errors that are not associated to errors in the measurement of the SHWFS frame, though it can be understood as dominated by the sampling of the wavefront through the number of subapertures. 
%he variance from the limited sampling of the wavefront phase due to the number of subapertures, and $\sigma_{\mathrm{fit}}^2$ is the variance from the limited number of modes used to reconstruct the wavefront phase. Note that there is no formal definition of what $\sigma_{\mathrm{fit}}^2$ is in the case of neural networks, so the fitting component and the aliasing component were not separated.
To measure  $\sigma_{\mathrm{reco}}^2$, the models were tested over the noise-free version of the testing datasets, computing the MSE of the residual phase. On the other hand, the $\sigma_{\mathrm{meas}}^2$ of each model was estimated by subtracting the outputs of the models in the case of noisy frames minus the outputs of the same frames without noise. The MSE was calculated over the resulting phase after the difference. The results of this analysis are shown in figure \ref{fig:errorbudget}.

Looking at figure \ref{fig:errorbudget}, it is evident that the best performing method is the ViT (blue), closely followed by the U-Net (orange). The MAP reconstructor is evaluated in two different regimes, analytical TWCoG (green) and CRLB (red) as mentioned in section \ref{sec:linear}. The ViT outperforms the U-Net in the reconstruction variance (x-axis) in all datasets and it also outperforms the U-Net in measurement variance (y-axis) in the case of $32\times32$ subapertures, although for the subsampled cases, the U-Net outperforms the ViT in measurement variance.

It can be argued that the attention mechanism of the ViT is more capable of modeling nonlinearities that are present in the SHWFS frames in these datasets, compared to the U-Net models. Since the U-Net uses convolutions, it can learn local features easier. That is, filtering noise within one subaperture and measuring spot shape. However, the U-Net presents a trade-off for less performance in high-frequency information within the entire pupil.

Nevertheless, both the ViT and U-Net models handle high-frequency features well when compared to linear reconstructors, considering the models are not designed taking into account atmosphere characteristics, unlike the MAP reconstructors which utilize KL-mode variance as priors. Furthermore, it is interesting to observe that deep-learning models trained exclusively in noisy data, yield a better performance in a dataset with no noise. This suggests that these models learn the underlying physical phenomena in some way, and properly separate between signal and noise.

The median variance over each KL mode of the residuals on the 32$\times$32 SHWFS testing dataset of each model is computed and shown in the left plot of figure \ref{fig:klvariance}. The ViT shows a significant reduction in variance over all KL modes compared to the U-Net and the MAP reconstructors. It is important to note how the deep-learning models keep an advantage in higher order modes much further than a 32$\times$32 SHWFS would suggest based on its sampling, making it evident that these models use more information than just spot position to make the reconstruction. 

In the right plot of figure \ref{fig:klvariance}, the variance present in each mode for each reconstructor is shown, where 0\% means a perfect reconstruction and 100\% no reconstruction at all. Note that the MAP reconstructors have a cut-off at 1000 modes, since this is the number of modes chosen for building the reconstruction matrices. Moreover, the ViT tends to be approximately 1.5 times better than the U-Net at reconstructing each mode, which is significant improvement in higher-order modes, suggesting a better performance in regimes of interest such as partial AO correction.

\section{Summary and future work}\label{sec:summary}
A dataset of 3.5 million simulated, uncorrelated turbulent phase-image pairs across four SHWFS sizes (8×8, 16$\times$16, 20$\times$20 and 32×32 subapertures) modeled on the Subaru Telescope pupil and ULTIMATE-START truth-WFS was generated. Attention-based models (ViTs) were compared to highly-used U-Nets and a linear MAP reconstructor. The MAP reconstructor was tested in two distinct noise regimes: using a Cramér-Rao lower bound covariance and a TWCoG analytical covariance.
The ViT achieved the best overall reconstruction performance, closely followed by the U-Net. The ViT's attention mechanism looks highly capable of modeling global non-linearities across the entire pupil, with an approximate $\sim1.5$ better median reconstruction performance for each mode than the U-Net in the case of $32\times32$ subapertures. In contrast, the U-Net's convolutional layers excel at localized feature extraction (e.g., subaperture noise filtering and spot shape measurement) but it trades this off for slightly lower accuracy in representing global high-frequency pupil details.

Future work will focus on extending this analysis to closed-loop telemetry and LTAO systems, based on ULTIMATE-START. For this, a hardware-aware dataset will be generated in the Subaru Telescope using the AO3k deformable mirror. Then, the dataset will be extended to include on-sky telemetry once ULTIMATE-START commences science operation. As it was shown that ViTs are good offline wavefront reconstructors, large-scale ViT unsupervised pre-training experiment will be carried out with this hardware-aware dataset. Furthermore, analysis to the register tokens, self-attention maps and cross-attention maps will be explored to understand the reasoning within the ViT model.

\section{Acknowledgements}\label{sec:acknowledgements}
This work was supported by JST SPRING, Japan Grant Number JPMJSP2104, and by the Grant-in-Aid for Scientific Research (grant numbers: JP26K00751). Numerical computations were in part carried out on the GPU system at the Center for Computational Astrophysics, National Astronomical Observatory of Japan.

% References
\bibliography{report} % bibliography data in report.bib

@ARTICLE{2022OExpr..30.2991P,
       author = {{Pou}, B. and {Ferreira}, F. and {Quinones}, E. and {Gratadour}, D. and {Martin}, M.},
        title = "{Adaptive optics control with multi-agent model-free reinforcement learning}",
      journal = {Optics Express},
         year = 2022,
        month = jan,
       volume = {30},
       number = {2},
        pages = {2991},
          doi = {10.1364/OE.444099},
       adsurl = {https://ui.adsabs.harvard.edu/abs/2022OExpr..30.2991P}
}

@ARTICLE{2023PASP..135k4501W,
       author = {{Wong}, Alison P. and {Norris}, Barnaby R.~M. and {Deo}, Vincent and {Tuthill}, Peter G. and {Scalzo}, Richard and {Sweeney}, David and {Ahn}, Kyohoon and {Lozi}, Julien and {Vievard}, S{\'e}bastien and {Guyon}, Olivier},
        title = "{Nonlinear Wave Front Reconstruction from a Pyramid Sensor using Neural Networks}",
      journal = {\pasp},
         year = 2023,
        month = nov,
       volume = {135},
       number = {1053},
          eid = {114501},
        pages = {114501},
          doi = {10.1088/1538-3873/acfdcb},
archivePrefix = {arXiv},
       eprint = {2311.02595},
 primaryClass = {astro-ph.IM},
       adsurl = {https://ui.adsabs.harvard.edu/abs/2023PASP..135k4501W}
}

@ARTICLE{2021JATIS...7a9001W,
       author = {{Wong}, Alison P. and {Norris}, Barnaby R.~M. and {Tuthill}, Peter G. and {Scalzo}, Richard and {Lozi}, Julien and {Vievard}, S{\'e}bastien and {Guyon}, Olivier},
        title = "{Predictive control for adaptive optics using neural networks}",
      journal = {Journal of Astronomical Telescopes, Instruments, and Systems},
         year = 2021,
        month = jan,
       volume = {7},
          eid = {019001},
        pages = {019001},
          doi = {10.1117/1.JATIS.7.1.019001},
       adsurl = {https://ui.adsabs.harvard.edu/abs/2021JATIS...7a9001W}
}

@ARTICLE{maxpool2d,
       author = {{Lecun}, Y. and {Bottou}, L. and {Bengio}, Y. and {Haffner}, P.},
        title = "{Gradient-based learning applied to document recognition}",
      journal = {IEEE Proceedings},
         year = 1998,
        month = nov,
       volume = {86},
       number = {11},
        pages = {2278-2324},
          doi = {10.1109/5.726791},
       adsurl = {https://ui.adsabs.harvard.edu/abs/1998IEEEP..86.2278L}
}

@ARTICLE{resunet,
       author = {{Zhang}, Zhengxin and {Liu}, Qingjie and {Wang}, Yunhong},
        title = "{Road Extraction by Deep Residual U-Net}",
      journal = {IEEE Geoscience and Remote Sensing Letters},
         year = 2018,
        month = may,
       volume = {15},
       number = {5},
        pages = {749-753},
          doi = {10.1109/LGRS.2018.2802944},
archivePrefix = {arXiv},
       eprint = {1711.10684},
 primaryClass = {cs.CV},
       adsurl = {https://ui.adsabs.harvard.edu/abs/2018IGRSL..15..749Z}
}

@ARTICLE{gelu,
       author = {{Hendrycks}, Dan and {Gimpel}, Kevin},
        title = "{Gaussian Error Linear Units (GELUs)}",
      journal = {arXiv e-prints},
         year = 2016,
        month = jun,
          eid = {arXiv:1606.08415},
        pages = {arXiv:1606.08415},
          doi = {10.48550/arXiv.1606.08415},
archivePrefix = {arXiv},
       eprint = {1606.08415},
 primaryClass = {cs.LG},
       adsurl = {https://ui.adsabs.harvard.edu/abs/2016arXiv160608415H}
}

@ARTICLE{groupnorm,
       author = {{Wu}, Yuxin and {He}, Kaiming},
        title = "{Group Normalization}",
      journal = {arXiv e-prints},
         year = 2018,
        month = mar,
          eid = {arXiv:1803.08494},
        pages = {arXiv:1803.08494},
          doi = {10.48550/arXiv.1803.08494},
archivePrefix = {arXiv},
       eprint = {1803.08494},
 primaryClass = {cs.CV},
       adsurl = {https://ui.adsabs.harvard.edu/abs/2018arXiv180308494W}
}

@INPROCEEDINGS{ultimate,
       author = {{Minowa}, Yosuke and {Koyama}, Yusei and {Ono}, Yoshito and {Tanaka}, Ichi and {Hattori}, Takashi and {Clergeon}, Christophe and {Akiyama}, Masayuki and {Kodama}, Tadayuki and {Motohara}, Kentaro and {Rigaut}, Fran{\c{c}}ois and et al.},
        title = "{ULTIMATE-Subaru: enhancing the Subaru's wide-field capability with GLAO}",
    booktitle = {Advances in Optical Astronomical Instrumentation 2019},
         year = 2020,
       editor = {{Ellis}, Simon C. and {d'Orgeville}, C{\'e}line},
       series = {Society of Photo-Optical Instrumentation Engineers (SPIE) Conference Series},
       volume = {11203},
        month = jan,
          eid = {112030G},
        pages = {112030G},
          doi = {10.1117/12.2560539},
       adsurl = {https://ui.adsabs.harvard.edu/abs/2020SPIE11203E..0GM}
}

@ARTICLE{landman2,
       author = {{Landman}, R. and {Haffert}, S.~Y. and {Long}, J.~D. and {Males}, J.~R. and {Close}, L.~M. and {Foster}, W.~B. and {Van Gorkom}, K. and {Guyon}, O. and {Hedglen}, A.~D. and {Johnson}, P.~T. and et al.},
        title = "{Making the unmodulated pyramid wavefront sensor smart: II. First on-sky demonstration of extreme adaptive optics with deep learning}",
      journal = {\aap},
         year = 2025,
        month = apr,
       volume = {696},
          eid = {L1},
        pages = {L1},
          doi = {10.1051/0004-6361/202553753},
archivePrefix = {arXiv},
       eprint = {2503.16690},
 primaryClass = {astro-ph.IM},
       adsurl = {https://ui.adsabs.harvard.edu/abs/2025A&A...696L...1L}
}

@ARTICLE{landman1,
       author = {{Landman}, R. and {Haffert}, S.~Y. and {Males}, J.~R. and {Close}, L.~M. and {Foster}, W.~B. and {Van Gorkom}, K. and {Guyon}, O. and {Hedglen}, A. and {Kautz}, M. and {Kueny}, J.~K. and et al.},
        title = "{Making the unmodulated Pyramid wavefront sensor smart. Closed-loop demonstration of neural network wavefront reconstruction with MagAO-X}",
      journal = {\aap},
         year = 2024,
        month = apr,
       volume = {684},
          eid = {A114},
        pages = {A114},
          doi = {10.1051/0004-6361/202348898},
archivePrefix = {arXiv},
       eprint = {2401.16325},
 primaryClass = {astro-ph.IM},
       adsurl = {https://ui.adsabs.harvard.edu/abs/2024A&A...684A.114L}
}

@ARTICLE{fp2,
       author = {{Orban de Xivry}, G. and {Quesnel}, M. and {Vanberg}, P.-O. and {Absil}, O. and {Louppe}, G.},
        title = "{Focal plane wavefront sensing using machine learning: performance of convolutional neural networks compared to fundamental limits}",
      journal = {\mnras},
         year = 2021,
        month = aug,
       volume = {505},
       number = {4},
        pages = {5702-5713},
          doi = {10.1093/mnras/stab1634},
archivePrefix = {arXiv},
       eprint = {2106.04456},
 primaryClass = {astro-ph.IM},
       adsurl = {https://ui.adsabs.harvard.edu/abs/2021MNRAS.505.5702O}
}

@INPROCEEDINGS{AO3k,
       author = {{Lozi}, Julien and {Ahn}, Kyohoon and {Blue}, Hannah and {Chun}, Alicia and {Clergeon}, Christophe and {Deo}, Vincent and {Guyon}, Olivier and {Hattori}, Takashi and {Minowa}, Yosuke and {Nishiyama}, Shogo and et al.},
        title = "{AO3k at Subaru: first on-sky results of the facility extreme-AO}",
    booktitle = {Adaptive Optics Systems IX},
         year = 2024,
       editor = {{Jackson}, Kathryn J. and {Schmidt}, Dirk and {Vernet}, Elise},
       series = {Society of Photo-Optical Instrumentation Engineers (SPIE) Conference Series},
       volume = {13097},
        month = aug,
          eid = {1309703},
        pages = {1309703},
          doi = {10.1117/12.3019294},
       adsurl = {https://ui.adsabs.harvard.edu/abs/2024SPIE13097E..03L}
}

@ARTICLE{jalofp,
       author = {{Nousiainen}, J. and {Taskin}, I. and {Kasper}, M. and {Orban de Xivry}, G. and {Absil}, O.},
        title = "{Focal plane wavefront control with model-based reinforcement learning: I. Proof of concept on simulated static and dynamic non-common path aberrations}",
      journal = {\aap},
         year = 2026,
        month = may,
       volume = {709},
          eid = {A267},
        pages = {A267},
          doi = {10.1051/0004-6361/202558504},
archivePrefix = {arXiv},
       eprint = {2604.00993},
 primaryClass = {astro-ph.IM},
       adsurl = {https://ui.adsabs.harvard.edu/abs/2026A&A...709A.267N}
}

@ARTICLE{PO4AO3,
       author = {{Nousiainen}, Jalo and {Chambouleyron}, Vincent and {Neichel}, Benoit and {Cetre}, Sylvain and {Sauvage}, Jean-Francois and {Alagao}, Angelie and {Kasper}, Markus and {Dray}, Jonathan and {Fetick}, Romain and {Engler}, Byron},
        title = "{On-sky demonstration of reinforcement learning for adaptive optics control}",
      journal = {arXiv e-prints},
         year = 2026,
        month = jun,
          eid = {arXiv:2606.10771},
        pages = {arXiv:2606.10771},
          doi = {10.48550/arXiv.2606.10771},
archivePrefix = {arXiv},
       eprint = {2606.10771},
 primaryClass = {astro-ph.IM},
       adsurl = {https://ui.adsabs.harvard.edu/abs/2026arXiv260610771N}
}

@ARTICLE{PO4AO2,
       author = {{Nousiainen}, Jalo and {Engler}, Byron and {Kasper}, Markus and {Rajani}, Chang and {Helin}, Tapio and {Heritier}, C{\'e}dric T. and {Quanz}, Sascha P. and {Glauser}, Adrian M.},
        title = "{Laboratory experiments of model-based reinforcement learning for adaptive optics control}",
      journal = {Journal of Astronomical Telescopes, Instruments, and Systems},
         year = 2024,
        month = jan,
       volume = {10},
          eid = {019001},
        pages = {019001},
          doi = {10.1117/1.JATIS.10.1.019001},
archivePrefix = {arXiv},
       eprint = {2401.00242},
 primaryClass = {astro-ph.IM},
       adsurl = {https://ui.adsabs.harvard.edu/abs/2024JATIS..10a9001N}
}

@ARTICLE{PO4AO1,
       author = {{Nousiainen}, J. and {Rajani}, C. and {Kasper}, M. and {Helin}, T. and {Haffert}, S.~Y. and {V{\'e}rinaud}, C. and {Males}, J.~R. and {Van Gorkom}, K. and {Close}, L.~M. and {Long}, J.~D. and et al.},
        title = "{Toward on-sky adaptive optics control using reinforcement learning. Model-based policy optimization for adaptive optics}",
      journal = {\aap},
         year = 2022,
        month = aug,
       volume = {664},
          eid = {A71},
        pages = {A71},
          doi = {10.1051/0004-6361/202243311},
archivePrefix = {arXiv},
       eprint = {2205.07554},
 primaryClass = {astro-ph.IM},
       adsurl = {https://ui.adsabs.harvard.edu/abs/2022A&A...664A..71N}
}

@ARTICLE{2023JATIS...9a9001S,
       author = {{Smith}, Jeffrey and {Cranney}, Jesse and {Gretton}, Charles and {Gratadour}, Damien},
        title = "{Image-to-image translation for wavefront and point spread function estimation}",
      journal = {Journal of Astronomical Telescopes, Instruments, and Systems},
         year = 2023,
        month = jan,
       volume = {9},
          eid = {019001},
        pages = {019001},
          doi = {10.1117/1.JATIS.9.1.019001},
       adsurl = {https://ui.adsabs.harvard.edu/abs/2023JATIS...9a9001S}
}

@ARTICLE{rl1,
       author = {{Pou}, Bartomeu and {Smith}, Jeffrey and {Quinones}, Eduardo and {Martin}, Mario and {Gratadour}, Damien},
        title = "{Integrating supervised and reinforcement learning for predictive control with an unmodulated pyramid wavefront sensor for adaptive optics}",
      journal = {Optics Express},
         year = 2024,
        month = oct,
       volume = {32},
       number = {21},
        pages = {37011},
          doi = {10.1364/OE.530254},
archivePrefix = {arXiv},
       eprint = {2405.13610},
 primaryClass = {astro-ph.IM},
       adsurl = {https://ui.adsabs.harvard.edu/abs/2024OExpr..3237011P}
}

@ARTICLE{ViTinFP,
       author = {{Liu}, Xiaohan and {Luo}, Wen and {Hu}, Peng and {Zhang}, Jianzhu and {Zhang}, Feizhou and {Su}, Hua},
        title = "{Transformer-based wavefront sensing for atmospheric turbulence aberration correction}",
      journal = {\ao},
         year = 2025,
        month = apr,
       volume = {64},
       number = {10},
        pages = {2451},
          doi = {10.1364/AO.554333},
       adsurl = {https://ui.adsabs.harvard.edu/abs/2025ApOpt..64.2451L}
}

@ARTICLE{ViTinPyr,
       author = {{Weinberger}, Camilo and {Tapia}, Jorge and {Neichel}, Beno{\^\i}t and {Vera}, Esteban},
        title = "{Transformer neural networks for closed-loop adaptive optics using nonmodulated pyramid wavefront sensors}",
      journal = {\aap},
         year = 2024,
        month = jul,
       volume = {687},
          eid = {A202},
        pages = {A202},
          doi = {10.1051/0004-6361/202349118},
archivePrefix = {arXiv},
       eprint = {2405.05472},
 primaryClass = {astro-ph.IM},
       adsurl = {https://ui.adsabs.harvard.edu/abs/2024A&A...687A.202W}
}

@ARTICLE{ViT,
       author = {{Dosovitskiy}, Alexey and {Beyer}, Lucas and {Kolesnikov}, Alexander and {Weissenborn}, Dirk and {Zhai}, Xiaohua and {Unterthiner}, Thomas and {Dehghani}, Mostafa and {Minderer}, Matthias and {Heigold}, Georg and {Gelly}, Sylvain and {Uszkoreit}, Jakob and {Houlsby}, Neil},
        title = "{An Image is Worth 16x16 Words: Transformers for Image Recognition at Scale}",
      journal = {arXiv e-prints},
         year = 2020,
        month = oct,
          eid = {arXiv:2010.11929},
        pages = {arXiv:2010.11929},
          doi = {10.48550/arXiv.2010.11929},
archivePrefix = {arXiv},
       eprint = {2010.11929},
 primaryClass = {cs.CV},
       adsurl = {https://ui.adsabs.harvard.edu/abs/2020arXiv201011929D}
}

@ARTICLE{vaswani,
       author = {{Vaswani}, Ashish and {Shazeer}, Noam and {Parmar}, Niki and {Uszkoreit}, Jakob and {Jones}, Llion and {Gomez}, Aidan N. and {Kaiser}, Lukasz and {Polosukhin}, Illia},
        title = "{Attention Is All You Need}",
      journal = {arXiv e-prints},
         year = 2017,
        month = jun,
          eid = {arXiv:1706.03762},
        pages = {arXiv:1706.03762},
          doi = {10.48550/arXiv.1706.03762},
archivePrefix = {arXiv},
       eprint = {1706.03762},
 primaryClass = {cs.CL},
       adsurl = {https://ui.adsabs.harvard.edu/abs/2017arXiv170603762V}
}

@ARTICLE{U-Net,
       author = {{Ronneberger}, Olaf and {Fischer}, Philipp and {Brox}, Thomas},
        title = "{U-Net: Convolutional Networks for Biomedical Image Segmentation}",
      journal = {arXiv e-prints},
         year = 2015,
        month = may,
          eid = {arXiv:1505.04597},
        pages = {arXiv:1505.04597},
          doi = {10.48550/arXiv.1505.04597},
archivePrefix = {arXiv},
       eprint = {1505.04597},
 primaryClass = {cs.CV},
       adsurl = {https://ui.adsabs.harvard.edu/abs/2015arXiv150504597R}
}

@software{poppy,
       author = {{Perrin}, Marshall and {Long}, Joseph and {Douglas}, Ewan and {Sivaramakrishnan}, Anand and {Slocum}, Christine and {others}},
        title = "{POPPY: Physical Optics Propagation in PYthon}",
 howpublished = {Astrophysics Source Code Library, record ascl:1602.018},
         year = 2016,
        month = feb,
          eid = {ascl:1602.018},
archivePrefix = {ascl},
       eprint = {1602.018},
       adsurl = {https://ui.adsabs.harvard.edu/abs/2016ascl.soft02018P}
}

@ARTICLE{pytorch,
       author = {{Paszke}, Adam and {Gross}, Sam and {Massa}, Francisco and {Lerer}, Adam and {Bradbury}, James and {Chanan}, Gregory and {Killeen}, Trevor and {Lin}, Zeming and {Gimelshein}, Natalia and {Antiga}, Luca and {Desmaison}, Alban and {K{\"o}pf}, Andreas and {Yang}, Edward and {DeVito}, Zach and {Raison}, Martin and {Tejani}, Alykhan and {Chilamkurthy}, Sasank and {Steiner}, Benoit and {Fang}, Lu and {Bai}, Junjie and {Chintala}, Soumith},
        title = "{PyTorch: An Imperative Style, High-Performance Deep Learning Library}",
      journal = {arXiv e-prints},
         year = 2019,
        month = dec,
          eid = {arXiv:1912.01703},
        pages = {arXiv:1912.01703},
          doi = {10.48550/arXiv.1912.01703},
archivePrefix = {arXiv},
       eprint = {1912.01703},
 primaryClass = {cs.LG},
       adsurl = {https://ui.adsabs.harvard.edu/abs/2019arXiv191201703P}
}

@INPROCEEDINGS{start1,
       author = {{Akiyama}, Masayuki and {Minowa}, Yosuke and {Ono}, Yoshito and {Terao}, Koki and {Ogane}, Hajime and {Oomoto}, Kaoru and {Iizuka}, Yuta and {Oya}, Shin and {Mieda}, Etsuko and {Yamamuro}, Tomoyasu},
        title = "{ULTIMATE-START: Subaru tomography adaptive optics research experiment project overview}",
    booktitle = {Adaptive Optics Systems VII},
         year = 2020,
       editor = {{Schreiber}, Laura and {Schmidt}, Dirk and {Vernet}, Elise},
       series = {Society of Photo-Optical Instrumentation Engineers (SPIE) Conference Series},
       volume = {11448},
        month = dec,
          eid = {114481O},
        pages = {114481O},
          doi = {10.1117/12.2560872},
       adsurl = {https://ui.adsabs.harvard.edu/abs/2020SPIE11448E..1OA}
}

@INPROCEEDINGS{start2,
       author = {{Terao}, Koki and {Akiyama}, Masayuki and {Minowa}, Yosuke and {Ono}, Yoshito and {Ogane}, Hajime and {Akasawa}, Takumi and {Oya}, Shin and {Yamamuro}, Tomoyasu},
        title = "{ULTIMATE-START: current status of the Subaru Tomography Adaptive optics Research experimenT}",
    booktitle = {Adaptive Optics Systems VIII},
         year = 2022,
       editor = {{Schreiber}, Laura and {Schmidt}, Dirk and {Vernet}, Elise},
       series = {Society of Photo-Optical Instrumentation Engineers (SPIE) Conference Series},
       volume = {12185},
        month = aug,
          eid = {121856L},
        pages = {121856L},
          doi = {10.1117/12.2629652},
       adsurl = {https://ui.adsabs.harvard.edu/abs/2022SPIE12185E..6LT}
}

@INPROCEEDINGS{atmosphericparameters,
       author = {{Oya}, Shin and {Hayano}, Yutaka and {Lai}, Olivier and {Iwata}, Ikuru and {Kodama}, Tadayuki and {Arimoto}, Nobuo and {Minowa}, Yosuke and {Akiyama}, Masayuki and {Ono}, Yoshito H. and {Terada}, Hiroshi and {Usuda}, Tomonori and {Takami}, Hideki and {Nishimura}, Tetsuo and {Takato}, Naruhisa and {Tomono}, Daigo},
        title = "{ULTIMATE-SUBARU: simulation update}",
    booktitle = {Adaptive Optics Systems IV},
         year = 2014,
       editor = {{Marchetti}, Enrico and {Close}, Laird M. and {Vran}, Jean-Pierre},
       series = {Society of Photo-Optical Instrumentation Engineers (SPIE) Conference Series},
       volume = {9148},
        month = aug,
          eid = {91486G},
        pages = {91486G},
          doi = {10.1117/12.2055701},
       adsurl = {https://ui.adsabs.harvard.edu/abs/2014SPIE.9148E..6GO}
}

@INPROCEEDINGS{oopao,
       author = {{Taissir Heritier}, Cedric and {Verinaud}, Christophe and {Correia}, Carlos M.},
        title = "{OOPAO: Object Oriented Python Adaptive Optics}",
    booktitle = {Adaptive Optics for Extremely Large Telescopes (AO4ELT7)},
         year = 2023,
        month = jun,
          eid = {50},
        pages = {50},
          doi = {10.13009/AO4ELT7-2023-052},
       adsurl = {https://ui.adsabs.harvard.edu/abs/2023aoel.confE..50T}
}

@ARTICLE{adamw,
       author = {{Loshchilov}, Ilya and {Hutter}, Frank},
        title = "{Decoupled Weight Decay Regularization}",
      journal = {arXiv e-prints},
         year = 2017,
        month = nov,
          eid = {arXiv:1711.05101},
        pages = {arXiv:1711.05101},
          doi = {10.48550/arXiv.1711.05101},
archivePrefix = {arXiv},
       eprint = {1711.05101},
 primaryClass = {cs.LG},
       adsurl = {https://ui.adsabs.harvard.edu/abs/2017arXiv171105101L}
}

@ARTICLE{muon2,
       author = {{Liu}, Jingyuan and {Su}, Jianlin and {Yao}, Xingcheng and {Jiang}, Zhejun and {Lai}, Guokun and {Du}, Yulun and {Qin}, Yidao and {Xu}, Weixin and {Lu}, Enzhe and {Yan}, Junjie and {Chen}, Yanru and {Zheng}, Huabin and {Liu}, Yibo and {Liu}, Shaowei and {Yin}, Bohong and {He}, Weiran and {Zhu}, Han and {Wang}, Yuzhi and {Wang}, Jianzhou and {Dong}, Mengnan and {Zhang}, Zheng and {Kang}, Yongsheng and {Zhang}, Hao and {Xu}, Xinran and {Zhang}, Yutao and {Wu}, Yuxin and {Zhou}, Xinyu and {Yang}, Zhilin},
        title = "{Muon is Scalable for LLM Training}",
      journal = {arXiv e-prints},
         year = 2025,
        month = feb,
          eid = {arXiv:2502.16982},
        pages = {arXiv:2502.16982},
          doi = {10.48550/arXiv.2502.16982},
archivePrefix = {arXiv},
       eprint = {2502.16982},
 primaryClass = {cs.LG},
       adsurl = {https://ui.adsabs.harvard.edu/abs/2025arXiv250216982L}
}

@misc{muon1,
  author       = {Keller Jordan and Yuchen Jin and Vlado Boza and Jiacheng You and
                  Franz Cesista and Laker Newhouse and Jeremy Bernstein},
  title        = {Muon: An optimizer for hidden layers in neural networks},
  year         = {2024},
  url          = {https://kellerjordan.github.io/posts/muon/}
}

@ARTICLE{webdataset,
       author = {{Aizman}, Alex and {Maltby}, Gavin and {Breuel}, Thomas},
        title = "{High Performance I/O For Large Scale Deep Learning}",
      journal = {arXiv e-prints},
         year = 2020,
        month = jan,
          eid = {arXiv:2001.01858},
        pages = {arXiv:2001.01858},
          doi = {10.48550/arXiv.2001.01858},
archivePrefix = {arXiv},
       eprint = {2001.01858},
 primaryClass = {cs.DC},
       adsurl = {https://ui.adsabs.harvard.edu/abs/2020arXiv200101858A}
}

@ARTICLE{covcrlb,
       author = {{Winick}, Kim A.},
        title = "{Cram{\'e}r-Rao lower bounds on the performance of charge-coupled-device optical position estimators}",
      journal = {Journal of the Optical Society of America A},
         year = 1986,
        month = nov,
       volume = {3},
       number = {11},
        pages = {1809-1815},
          doi = {10.1364/JOSAA.3.001809},
       adsurl = {https://ui.adsabs.harvard.edu/abs/1986JOSAA...3.1809W}
}

@article{centroids,
    author = {Thomas, S. and Fusco, T. and Tokovinin, A. and Nicolle, M. and Michau, V. and Rousset, G.},
    title = {Comparison of centroid computation algorithms in a Shack–Hartmann sensor},
    journal = {Monthly Notices of the Royal Astronomical Society},
    volume = {371},
    number = {1},
    pages = {323-336},
    year = {2006},
    month = {09},
    issn = {0035-8711},
    doi = {10.1111/j.1365-2966.2006.10661.x},
    url = {https://doi.org/10.1111/j.1365-2966.2006.10661.x},
    eprint = {https://academic.oup.com/mnras/article-pdf/371/1/323/3029534/mnras0371-0323.pdf},
}

@article{southwell,
author = {W.H. Southwell},
journal = {J. Opt. Soc. Am.},
number = {8},
pages = {998--1006},
publisher = {Optica Publishing Group},
title = {Wave-front estimation from wave-front slope measurements},
volume = {70},
month = {Aug},
year = {1980},
url = {https://opg.optica.org/abstract.cfm?URI=josa-70-8-998},
doi = {10.1364/JOSA.70.000998},
}

@incollection{welch2025welch,
  title={The Perceptron},
  booktitle={The Welch Labs Illustrated Guide to AI},
  author={Welch, S. and Baskin, S. and Gundu, P.},
  isbn={9798991923415},
  chapter   = {1},
  url={https://books.google.co.jp/books?id=W8yb0QEACAAJ},
  year={2025},
  publisher={Welch Labs LLC}
}

@ARTICLE{attentionsink,
       author = {{Xiao}, Guangxuan and {Tian}, Yuandong and {Chen}, Beidi and {Han}, Song and {Lewis}, Mike},
        title = "{Efficient Streaming Language Models with Attention Sinks}",
      journal = {arXiv e-prints},
         year = 2023,
        month = sep,
          eid = {arXiv:2309.17453},
        pages = {arXiv:2309.17453},
          doi = {10.48550/arXiv.2309.17453},
archivePrefix = {arXiv},
       eprint = {2309.17453},
 primaryClass = {cs.CL},
       adsurl = {https://ui.adsabs.harvard.edu/abs/2023arXiv230917453X}
}

@ARTICLE{vit-5,
       author = {{Wang}, Feng and {Ren}, Sucheng and {Zhang}, Tiezheng and {Neskovic}, Predrag and {Bhattad}, Anand and {Xie}, Cihang and {Yuille}, Alan},
        title = "{ViT-5: Vision Transformers for The Mid-2020s}",
      journal = {arXiv e-prints},
         year = 2026,
        month = feb,
          eid = {arXiv:2602.08071},
        pages = {arXiv:2602.08071},
          doi = {10.48550/arXiv.2602.08071},
archivePrefix = {arXiv},
       eprint = {2602.08071},
 primaryClass = {cs.CV},
       adsurl = {https://ui.adsabs.harvard.edu/abs/2026arXiv260208071W}
}

@ARTICLE{qknorm,
       author = {{Henry}, Alex and {Dachapally}, Prudhvi Raj and {Pawar}, Shubham and {Chen}, Yuxuan},
        title = "{Query-Key Normalization for Transformers}",
      journal = {arXiv e-prints},
         year = 2020,
        month = oct,
          eid = {arXiv:2010.04245},
        pages = {arXiv:2010.04245},
          doi = {10.48550/arXiv.2010.04245},
archivePrefix = {arXiv},
       eprint = {2010.04245},
 primaryClass = {cs.CL},
       adsurl = {https://ui.adsabs.harvard.edu/abs/2020arXiv201004245H}
}

@ARTICLE{registers,
       author = {{Darcet}, Timoth{\'e}e and {Oquab}, Maxime and {Mairal}, Julien and {Bojanowski}, Piotr},
        title = "{Vision Transformers Need Registers}",
      journal = {arXiv e-prints},
         year = 2023,
        month = sep,
          eid = {arXiv:2309.16588},
        pages = {arXiv:2309.16588},
          doi = {10.48550/arXiv.2309.16588},
archivePrefix = {arXiv},
       eprint = {2309.16588},
 primaryClass = {cs.CV},
       adsurl = {https://ui.adsabs.harvard.edu/abs/2023arXiv230916588D}
}

@ARTICLE{rope,
       author = {{Su}, Jianlin and {Lu}, Yu and {Pan}, Shengfeng and {Murtadha}, Ahmed and {Wen}, Bo and {Liu}, Yunfeng},
        title = "{RoFormer: Enhanced Transformer with Rotary Position Embedding}",
      journal = {arXiv e-prints},
         year = 2021,
        month = apr,
          eid = {arXiv:2104.09864},
        pages = {arXiv:2104.09864},
          doi = {10.48550/arXiv.2104.09864},
archivePrefix = {arXiv},
       eprint = {2104.09864},
 primaryClass = {cs.CL},
       adsurl = {https://ui.adsabs.harvard.edu/abs/2021arXiv210409864S}
}

@ARTICLE{rmsnorm,
       author = {{Zhang}, Biao and {Sennrich}, Rico},
        title = "{Root Mean Square Layer Normalization}",
      journal = {arXiv e-prints},
         year = 2019,
        month = oct,
          eid = {arXiv:1910.07467},
        pages = {arXiv:1910.07467},
          doi = {10.48550/arXiv.1910.07467},
archivePrefix = {arXiv},
       eprint = {1910.07467},
 primaryClass = {cs.LG},
       adsurl = {https://ui.adsabs.harvard.edu/abs/2019arXiv191007467Z}
}

@INPROCEEDINGS{layerscale,
       author = {{Touvron}, Hugo and {Cord}, Matthieu and {Sablayrolles}, Alexandre and {Synnaeve}, Gabriel and {Jegou}, Herve},
        title = "{Going deeper with Image Transformers}",
    booktitle = {2021 IEEE/CVF International Conference on Computer Vision (ICCV)},
         year = 2021,
        month = oct,
    publisher = {IEEE},
          eid = {10},
        pages = {10},
          doi = {10.1109/ICCV48922.2021.00010},
       adsurl = {https://ui.adsabs.harvard.edu/abs/2021iccv.conf...10T}
}

@ARTICLE{subpixelconv,
       author = {{Shi}, Wenzhe and {Caballero}, Jose and {Husz{\'a}r}, Ferenc and {Totz}, Johannes and {Aitken}, Andrew P. and {Bishop}, Rob and {Rueckert}, Daniel and {Wang}, Zehan},
        title = "{Real-Time Single Image and Video Super-Resolution Using an Efficient Sub-Pixel Convolutional Neural Network}",
      journal = {arXiv e-prints},
         year = 2016,
        month = sep,
          eid = {arXiv:1609.05158},
        pages = {arXiv:1609.05158},
          doi = {10.48550/arXiv.1609.05158},
archivePrefix = {arXiv},
       eprint = {1609.05158},
 primaryClass = {cs.CV},
       adsurl = {https://ui.adsabs.harvard.edu/abs/2016arXiv160905158S}
}

@article{sobol2,
title = {Scrambling Sobol' and Niederreiter–Xing Points},
journal = {Journal of Complexity},
volume = {14},
number = {4},
pages = {466-489},
year = {1998},
issn = {0885-064X},
doi = {https://doi.org/10.1006/jcom.1998.0487},
url = {https://www.sciencedirect.com/science/article/pii/S0885064X98904873},
author = {Art B. Owen}
}

@article{sobol1,
title = {On the distribution of points in a cube and the approximate evaluation of integrals},
journal = {USSR Computational Mathematics and Mathematical Physics},
volume = {7},
number = {4},
pages = {86-112},
year = {1967},
issn = {0041-5553},
doi = {https://doi.org/10.1016/0041-5553(67)90144-9},
url = {https://www.sciencedirect.com/science/article/pii/0041555367901449},
author = {I.M Sobol'}
}

@article{sobol4,
author = {Joe, Stephen and Kuo, Frances Y.},
title = {Constructing Sobol Sequences with Better Two-Dimensional Projections},
journal = {SIAM Journal on Scientific Computing},
volume = {30},
number = {5},
pages = {2635-2654},
year = {2008},
doi = {10.1137/070709359},

URL = { 
    
        https://doi.org/10.1137/070709359
    
    

},
eprint = { 
    
        https://doi.org/10.1137/070709359
    
    

}
}

@article{sobol3,
author = {Joe, Stephen and Kuo, Frances Y.},
title = {Remark on algorithm 659: Implementing Sobol's quasirandom sequence generator},
year = {2003},
issue_date = {March 2003},
publisher = {Association for Computing Machinery},
address = {New York, NY, USA},
volume = {29},
number = {1},
issn = {0098-3500},
url = {https://doi.org/10.1145/641876.641879},
doi = {10.1145/641876.641879},
journal = {ACM Trans. Math. Softw.},
month = mar,
pages = {49–57},
numpages = {9}
}
\bibliographystyle{spiebib} % makes bibtex use spiebib.bst

\end{document}